# Metal-insulator transition induced in SrTi$_{1-x}$V$_x$O$_3$ thin films

Man Gu[1, a)], Stuart A. Wolf[1, 2] and Jiwei Lu[2, b)]

[1]Department of Physics, University of Virginia, 382 McCormick Rd., Charlottesville, VA 22904

[2]Department of Materials Science and Engineering, University of Virginia, 395 McCormick Rd.,

Charlottesville, VA 22904

Author to whom correspondence should be addressed:

[a)] E-mail: mg8yd@virginia.edu

[b)] E-mail: jl5tk@virginia.edu

**Abstract:**

Epitaxial SrTi$_{1-x}$V$_x$O$_3$ ($0 \leq x \leq 1$) thin films with thicknesses of ~16 nm were grown on (001)-oriented LSAT substrates using the pulsed electron-beam deposition technique. The transport study revealed a temperature driven metal-insulator transition (MIT) at 95 K for the film with $x = 0.67$. The films with higher vanadium concentration ($x > 0.67$) were metallic, and the electrical resistivity followed the $T^2$ law corresponding to a Fermi liquid system. In the insulating region of $x < 0.67$, the temperature dependence of electrical resistivity for the $x = 0.5$ and 0.33 films can be scaled with Mott's variable range hopping model. The possible mechanisms behind the observed MIT were discussed, including the effects of electron correlation, lattice distortion and Anderson localization.

Perovskite-type transition metal oxides $ABO_3$ represent one of the most important classes of functional materials, the strong interactions between spin, charge, orbital and lattice lead to a wide variety of intriguing electrical and magnetic properties. Strongly correlated oxides undergo a metal-insulator transition (MIT) are of particular interest, experimental and theoretical efforts to understand the MIT haven been going on for decades,[1] the underlying physical mechanisms include electron-electron interactions (Mott transition[2,3]), electron-phonon interactions (Peierls transition[4]) and disorder-induced localization (Anderson localization[5]). In recent years, major advances in thin film technology have provided great opportunities to explore nanoscale device applications utilizing the MIT in strongly correlated oxides,[6] and also opened new perspectives on the fundamental physics behind this fascinating phenomenon.

$SrVO_3$ (SVO) with a $3d^1$ electronic configuration for vanadium is a typical strongly correlated system, it has been found to exhibit metallic behavior with electrical resistivity ranging from $10^{-5}$ to $10^{-3}$ cm at room temperature.[7,8,9,10] According to the Hubbard model, a MIT can be induced by modifying the bandwidth or the band filling.[1] Bandwidth-controlled MIT (BC-MIT) has been observed in both SVO[11,12,13,14] and $CaVO_3$ (CVO)[15] ultrathin films and also on the surface of single-crystal SVO and CVO with thicknesses of a few nm.[16,17] Filling-controlled MIT (FC-MIT) has been intensively studied via aliovalent A-site substitution, such as in the $La_{1-x}Sr_xVO_3$ system.[18,19,20] Although few experimental studies have been done to study the MIT in SVO via B-site substitution, $SrTi_{1-x}V_xO_3$ (STVO), the solid solution between a correlated metal SVO ($3d^1$) and a band insulator $SrTiO_3$ (STO) ($3d^0$), offers a very interesting system to investigate the composition dependent MIT. Both SVO and STO are cubic perovskites with lattice constants close to each other ($a$ = 3.843 Å for SVO[8]; $a$ = 3.905 Å for STO). Since the element vanadium is next to titanium in the periodic table, the solid solution is expected to form over a wide range of compositions. In fact, STVO has been studied in the bulk, and MITs were observed at $x$ = 0.6-0.7.[21,22] However, the driving force behind the observed MIT is still under debate, Tsuiki *et al.* interpreted their findings by assuming the local Jahn-Teller distortion around $V^{4+}$ ion,[21] and Hong *et al.* proposed a model of Anderson localized states in which a MIT occurs where the mobility edge crosses the Fermi level.[22] In this work, we further explore this solid solution system by investigating the transport properties of the STVO thin films over the whole composition range of $0 \leq x \leq 1$, and discuss the possible mechanisms behind the observed MIT.

High-quality epitaxial STVO thin films with thicknesses of ~16 nm were deposited on (001)-oriented $(LaAlO_3)_{0.3}(Sr_2AlTaO_6)_{0.7}$ (LSAT) ($a$ = 3.868 Å) substrates using a pulsed electron-beam deposition (PED) (Neocera Inc.) technique.[23] The base pressure of the vacuum chamber was ~$5 \times 10^{-8}$ Torr. The PED system is equipped with two electron-beam guns. The STVO thin films were co-deposited from two individual ceramic targets. In the case of SVO, a 12 kV potential in the electron-beam gun was applied to ablate a $Sr_2V_2O_7$ target, the deposition rate was found to be 0.006 Å/pulse which equates to 1.8 Å/min at a 5 Hz pulse rate. For STO, a stoichiometric STO target was used and the electron-beam gun was operated at 10 kV to achieve a similar growth rate as SVO. The composition of the STVO thin films was controlled by varying the pulse rate for each of the two targets in the range of 1-5 Hz. All the films in this study were grown at a substrate temperature of 800 °C in a 10 mTorr Ar atmosphere.

The film surface morphology was analyzed with atomic force microscopy (AFM) (Cypher, Asylum Research Inc.), the growth of the STVO thin films on LSAT (001) substrates was carried out in a layer-by-layer growth mode, and all the films showed atomically flat surfaces with the RMS roughness of ~0.2 nm. The film crystalline structure was examined by x-ray diffraction (XRD) (Smartlab, Rigaku Inc.) using Cu K$\alpha$ radiation. X-ray reflectivity (XRR) measurements were performed to determine the film thickness, and all the films showed a similar thickness of ~16 nm. The electrical resistivity and Hall effects of the films were measured in the van der Pauw geometry with cold-welded indium contacts using a Physical Property Measurement System (PPMS) (Quantum Design Inc.) in the temperature range of 2-300 K. After the measurements, Ti(10 nm)/Au(100 nm) contacts were deposited on the films by electron-beam evaporation to study the magnetoresistance (MR) using the PPMS at low temperature between 2 K to 30 K.

As shown in Fig. 1a, out-of-plane XRD scans confirmed single phase STVO solid solution system in the whole composition range of $0 \leq x \leq 1$, the Kiessig fringes around (001) and (002) diffraction peaks indicated coherent films with smooth interfaces. Since SVO has a smaller lattice parameter than STO, the STVO film peaks gradually shifted to the higher $2\theta$ value with the increase of the vanadium concentration $x$, the calculated out-of-plane lattice parameters $c$ are listed in TABLE I. The reciprocal space mapping on the $(\bar{1}03)$ asymmetric reflection in Figs. 1c-i also confirmed coherent growth of all the STVO ($0 \leq x \leq 1$) thin films. The in-plane lattice parameters $a$ were found to be 3.868 Å for all the films, since the films were coherently strained to the in-plane lattice parameter of the LSAT substrates. As listed in TABLE I, the unit cell volume $V$ were calculated from both out-of-plane and in-plane lattice parameters. Following Vegard's law, $V$ is an approximately linear function of $x$ as plotted in Fig. 1b.

Fig. 2a shows the electrical resistivity as a function of temperature for the STVO ($0.33 \leq x \leq 1$) thin films in the temperature range of 2-300 K (the $x = 0.17$ and 0 films were too insulating for any transport measurements), the resistivity data is listed in TABLE II. Reducing the vanadium concentration $x$ caused an increase in film resistivity over the entire temperature range, and a transition from a metallic to a semiconducting temperature dependence with the $x = 0.67$ film at the boundary showing a temperature driven MIT. The films with higher vanadium concentration ($x = 0.83$ and 1) were metallic, and the ones with lower vanadium concentration ($x = 0.33$ and 0.5) were semiconducting. Also, room-temperature Hall effects were investigated for the STVO ($0.5 \leq x \leq 1$) thin films. The carriers were found to be electrons, and the carrier density decreased with the decrease of $x$ as shown in Fig. 2b.

The detailed temperature dependence of the transport properties for the STVO ($0.33 \leq x \leq 1$) thin films are shown in Figs. 3a-d. Both $x = 1$ and $x = 0.83$ films exhibited metallic behaviors (Fig. 3a), the electrical resistivity of the $x = 0.83$ film was slightly higher than that of the pure SVO film ($x = 1$). Over the entire temperature range of 2-300 K, the electrical resistivity as a function of temperature followed the $\rho = \rho_0 + AT^2$ relationship corresponding to a Fermi liquid model, in which the residual resistivity $\rho_0$ is a temperature independent value from the electron-impurity scattering caused by defects, and $A$ quantifies the electron-electron interactions. In Fig. 3b, the $x = 0.67$ film exhibited a temperature driven MIT characterized by a resistivity upturn at a transition temperature $T_{MIT}$ of 95 K (indicated by the arrow). Above $T_{MIT}$, the $x = 0.67$ film showed metallic behavior with the temperature dependence of resistivity still following the $T^2$ law. As listed in TABLE II, an increase in the estimated values of $\rho_0$ was observed with decreasing $x$, and the $x = 0.67$ film showed a much higher $A$ value than the $x = 0.83$ and 1 films. As the temperature was lowered from $T_{MIT}$, the resistivity of $x = 0.67$ film gradually rose up, and then dropped a little below 35 K. The temperature dependence of conductivity between 45 K and 90 K followed $\exp[-(T_0/T)^{1/4}]$ law consistent with Mott's variable range hopping (VRH) model (inset of Fig. 3b). Also, as plotted in Fig. 3c and 3d, the semiconducting $x = 0.5$ and 0.33 films all showed good fits to the VRH model in the entire temperature range, a dramatic increase in the fitting parameter $T_0$ with decreasing $x$ was observed (TABLE II).

Our observation of the MIT induced in the STVO ($x = 0.67$) thin film is in good agreement with the literature data for the bulk material.[21,22] Tsuiki et al.[21] reported a MIT in the $x \approx 0.6$ sample and VRH mechanisms for the samples with lower vanadium concentration ($x$ 0.6). The high critical donor concentration $x$ for the MIT was interpreted by a very deep donor level by assuming a Jahn-Teller distortion induced around $V^{4+}$ ion in the STO lattice, the local Jahn-Teller distortion makes the electrons self-trapped and strongly localized to the donor center. On the other hand, Hong et al.[22] observed a temperature driven MIT in the $x = 0.7$ sample with a transition temperature $T_{MIT}$ of 120 K, which is quite similar to our result. However, a different conduction mechanism in the semiconducting region was observed, the low temperature data of the $x = 0.7$ sample can be scaled with a 3D weak localization model, and a model of Anderson localized states was proposed in which a MIT occurs where the mobility edge crosses the Fermi level. Moreover, the study of a similar system of $CaV_{1-x}Ti_xO_3$ revealed that the substitution of $Ti^{4+}$ for $V^{4+}$ ions may also transform the strong-correlation fluctuations by narrowing the * band in addition to introducing Anderson localized states.[24]

It is well-known that electron-electron interactions play an important role in the Anderson-localized regime of the impurity band in doped semiconductors.[3] Low temperature magneto-transport study was carried out to understand the transport mechanism in the STVO thin films. As shown in Fig. 4, MR (defined as $[R(B)–R(0)]/R(0)$) in a perpendicular magnetic field was measured for the $x = 0.67$ film in the low temperature of 2-30 K. A small positive MR was observed for all the temperatures, the MR increased with the decrease in temperature, and at each temperature point, the MR was found to be roughly proportional to $B^2$. Our observations are not expected from Anderson localization, since the magnetic field suppresses the coherent interference for localization and results in a negative MR.[25,26] In fact, the effects of electron-electron interactions on the MR in the Anderson-localized regime have been investigated, in the presence of a magnetic field, the interplay of the intrastate interaction and the energy dependence of localization lengths results in the competition of positive and negative contributions in the field dependence of the MR.[27,28] Therefore, the positive MR observed in the STVO thin films may suggest that the effect of electron-electron interactions is not negligible, and Anderson localization is not the only driving force behind the observed MIT.

A possible interpretation of the MIT induced in the STVO thin films can rely on the interplay between electron-electron interactions and disorder-induced localization. In the STVO system with $Ti^{4+}$ ($3d^0$) and $V^{4+}$ ($3d^1$) ions randomly occupying the B sites, the $Ti^{4+}$ ions perturb the periodic potential of the V $3d$ band so as to introduce significant disorder and Anderson-localized states. In addition, lattice distortions caused by the epitaxial strain may also have dramatic consequences on the transport properties of the STVO thin films. As obtained from the film crystalline structure, all the STVO films had a same in-plane lattice parameter $a$ as the films were completely strained to the LSAT substrates, and the out-of-plane lattice parameter $c$ was found to increase with decreasing $x$. The $c/a$ ratio is listed in TABLE I, the lattice structure gradually transformed from horizontally stretched to compressed as $x$ decreases, and the $x = 0.67$ film was closer to a cubic lattice. The lattice distortion with decreasing $x$ causes a greater separation between the adjacent $V^{4+}$ ions and thus less overlap between the $3d$ orbital states, which may result in a narrower V $3d$ band. According to the Hubbard model, the MIT in a strongly correlated system can be controlled by the $U/W$ ratio, where $U$ is the on-site Coulomb repulsion (localizes the electron), and $W$ is the one-electron bandwidth (the tendency of electrons to delocalize), a MIT occurs when the ratio is beyond a critical value.[1] Thus, the decrease in the V concentration $x$ results in a reduction of the effective bandwidth $W$, and the films undergo a composition dependent MIT when the V concentration crosses $x = 0.67$. Since the $x = 0.67$ film lies at the boundary of the metallic and insulating films, the observed more complicated temperature dependence may result from a transition region of a spatial inhomogeneity with both metallic and insulating phases coexisting, which can be thought of as a percolative behavior.[29,30]

In conclusion, we have successfully synthesized high-quality epitaxial STVO ($0 \leq x \leq 1$) thin films on LSAT substrates. The transport properties were found to be strongly dependent on the film composition. A temperature driven MIT was observed in the $x = 0.67$ film at 95 K. The films with higher vanadium concentration ($x = 0.83$ and 1) were metallic, and the ones with lower vanadium concentration ($x = 0.33$ and 0.5) were semiconducting. The possible mechanism behind the observed MIT might be associated the interplay between electron-electron interactions and disorder-induced localization. The $Ti^{4+}$ ion substitution introduces Anderson-localized states as well as lattice distortions that result in a reduction in the effective $3d$ bandwidth $W$. STVO offers a very interesting system to investigate the effects of electron-electron interactions in the Anderson-localized regime, the film electronic structures will be studied in our future work.

**Acknowledgements:**

We gratefully acknowledge the financial support from the Army Research Office through MURI grant No. W911-NF-09-1-0398. The authors would like to thank Drs. Ryan Comes, Hongxue Liu and Mr. Yonghang Pei for their helpful discussions.

**References:**

[1] M. Imada, A. Fujimori, and Y. Tokura, Rev Mod Phys **70**, 1039 (1998).

[2] N. F. Mott, Proc. Phys. Soc. London, Sect. A **62**, 416 (1949).

[3]   N. F. Mott, *Metal-Insulator Transitions*, 2nd ed. (Taylor and Francis, London, 1990).

[4]   R. E. Peierls, *Quantum Theory of Solids*. (Oxford University Press, New York, 1955).

[5]   P. W. Anderson,  Phys. Rev. **109**, 1492 (1958).

[6]   Z. Yang, C. Y. Ko, and S. Ramanathan,  Annu Rev Mater Res **41**, 337 (2011).

[7]   B. L. Chamberland and P. S. Danielson,  J Solid State Chem **3**, 243 (1971).

[8]   P. Dougier, J. C. C. Fan, and J. B. Goodenough,  J Solid State Chem **14**, 247 (1975).

[9]   M. Onoda, H. Ohta, and H. Nagasawa,  Solid State Commun **79**, 281 (1991).

[10]  V. Giannakopoulou, P. Odier, J. M. Bassat, and J. P. Loup,  Solid State Commun **93**, 579 (1995).

[11]  K. Yoshimatsu, T. Okabe, H. Kumigashira, S. Okamoto, S. Aizaki, A. Fujimori, and M. Oshima,  Phys Rev Lett **104**, 147601 (2010).

[12]  K. Yoshimatsu, K. Horiba, H. Kumigashira, T. Yoshida, A. Fujimori, and M. Oshima,  Science **333**, 319 (2011).

[13]  S. Okamoto,  Phys Rev B **84**, 201305(R) (2011).

[14]  M. Gu, Stuart A. Wolf, and J. W. Lu,  arXiv:1307.5819 [cond-mat.str-el] (2013).

[15]  M. Gu, J. Laverock, B. Chen, K. E. Smith, S. A. Wolf, and J. W. Lu,  J Appl Phys **113**, 133704 (2013).

[16]  K. Maiti, D. D. Sarma, M. J. Rozenberg, I. H. Inoue, H. Makino, O. Goto, M. Pedio, and R. Cimino,  Europhys Lett **55**, 246 (2001).

[17]  J. Laverock, B. Chen, K. E. Smith, R. P. Singh, G. Balakrishnan, M. Gu, J. W. Lu, S. A. Wolf, R. M. Qiao, W. Yang, and J. Adell,  Phys Rev Lett **111**, 047402 (2013).

[18]  F. Inaba, T. Arima, T. Ishikawa, T. Katsufuji, and Y. Tokura,  Phys Rev B **52**, R2221 (1995).

19  S. Miyasaka, T. Okuda, and Y. Tokura, Phys Rev Lett **85**, 5388 (2000).

20  T. M. Dao, P. S. Mondal, Y. Takamura, E. Arenholz, and J. Lee, Appl Phys Lett **99**, 112111 (2011).

21  H. Tsuiki, K. Kitazawa, and K. Fueki, Jpn J Appl Phys **22**, 590 (1983).

22  K. Y. Hong, S. H. Kim, Y. J. Heo, and Y. U. Kwon, Solid State Commun **123**, 305 (2002).

23  R. Comes, M. Gu, M. Khokhlov, H. X. Liu, J. W. Lu, and S. A. Wolf, J Appl Phys **113**, 023303 (2013).

24  H. D. Zhou and J. B. Goodenough, Phys Rev B **69**, 245118 (2004).

25  G. Bergmann, Phys Rep **107**, 1 (1984).

26  B. L. Altshuler, D. Khmel'nitzkii, A. I. Larkin, and P. A. Lee, Phys Rev B **22**, 5142 (1980).

27  H. Kamimura, Supplement of the Progress of Theoretical Physics **72**, 206 (1982).

28  A. Kurobe, J Phys C Solid State **19**, 2201 (1986).

29  O. Entin-Wohlman, Y. Gefen, and Y. Shapira, J Phys C Solid State **16**, 1161 (1983).

30  A. M. Goldman and S. A. Wolf, *Percolation, Localization, and Superconductivity*. (Plenum Press, New York and London, 1984).

TABLE I. Lattice parameters of the STVO (0 ≤ $x$ ≤ 1) thin films. (For all the films, in-plane lattice parameters $a$ = 3.868 Å).

| V concentration, $x$ | Out-of-plane lattice parameter, $c$ (Å) | $c/a$ | Unit cell volume, $V$ (Å$^3$) |
|---|---|---|---|
| 1 | 3.829 | 0.990 | 57.287 |
| 0.83 | 3.847 | 0.995 | 57.557 |
| 0.67 | 3.879 | 1.003 | 58.035 |
| 0.5 | 3.926 | 1.015 | 58.739 |
| 0.33 | 3.973 | 1.027 | 59.442 |
| 0.17 | 4.013 | 1.037 | 60.040 |
| 0 | 4.023 | 1.040 | 60.190 |

TABLE II. Transport properties of the STVO (0.33 ≤ $x$ ≤ 1) thin films.

| V concentration, $x$ | $\rho$(300 K) (Ω cm) | $\rho$(2 K) (Ω cm) | $\rho_0$ (Ω cm) | $A$ (Ω cm/K$^2$) | $T_0$ (K) |
|---|---|---|---|---|---|
| 1 | $2.30 \times 10^{-4}$ | $1.52 \times 10^{-4}$ | $1.51 \times 10^{-4}$ | $8.79 \times 10^{-10}$ | - |
| 0.83 | $2.33 \times 10^{-4}$ | $1.64 \times 10^{-4}$ | $1.65 \times 10^{-4}$ | $7.54 \times 10^{-10}$ | - |
| 0.67 | $5.73 \times 10^{-4}$ | $4.60 \times 10^{-4}$ | $4.31 \times 10^{-4}$ | $1.57 \times 10^{-9}$ | $4.24 \times 10^{-4}$ |
| 0.5 | $6.65 \times 10^{-3}$ | 3.40 | - | - | $1.16 \times 10^{4}$ |
| 0.33 | 0.47 | > 20 | - | - | $1.07 \times 10^{7}$ |

**Figure Captions:**

Fig. 1

(a) XRD scans for the STVO ($0 \leq x \leq 1$) thin films grown on LSAT (001) substrates. (b) Calculated unit cell volume as a function of $x$. The straight line represents the Vegard's law. (c)-(i) The reciprocal space mapping on $(\bar{1}03)$ reflection for the same STVO thin films.

Fig. 2

(a) Electrical resistivity vs. temperature for the STVO ($0.33 \leq x \leq 1$) thin films. (b) Carrier density from room-temperature Hall measurements for the STVO ($0.5 \leq x \leq 1$) thin films.

Fig. 3

Temperature dependence of electrical resistivity for the STVO thin films, (a) $x = 1$ and 0.83, (b) $x = 0.67$, $\rho = \rho_0 + AT^2$ fits are also shown. Logarithm of conductivity as a function of $T^{1/4}$ with a linear fit to the data, (inset of b) $x = 0.67$, 45 K $\leq T \leq$ 90 K, (c) $x = 0.5$, 2 K $\leq T \leq$ 300 K, (d) $x = 0.33$, 110 K $\leq T \leq$ 300 K.

Fig. 4

Normalized out-of-plane MR of the STVO ($x = 0.67$) thin film between 2 K and 30 K with the field up to 7 T. The parabolic fits at each temperature are also shown in lines.

Fig. 1

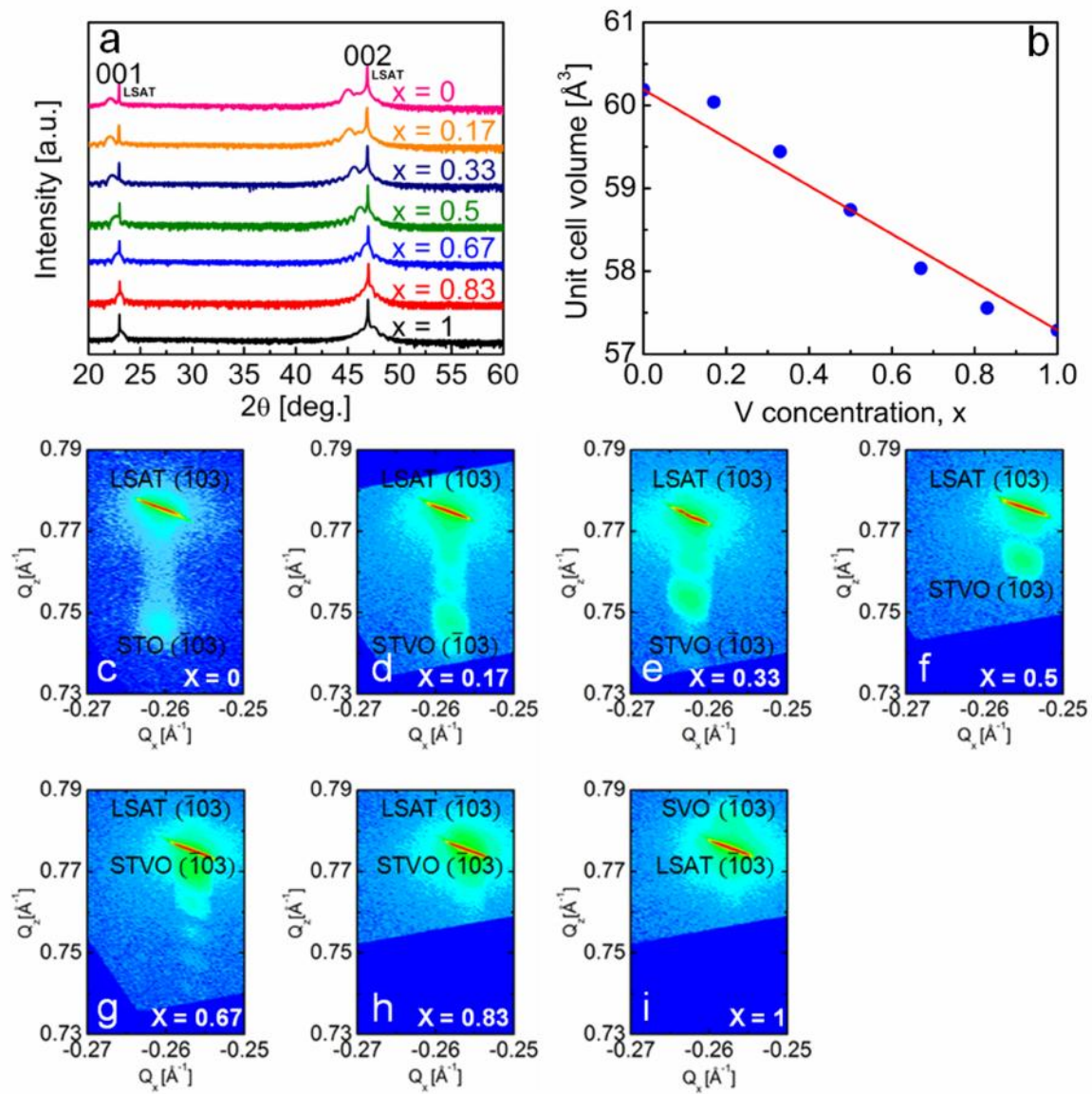

Fig. 2

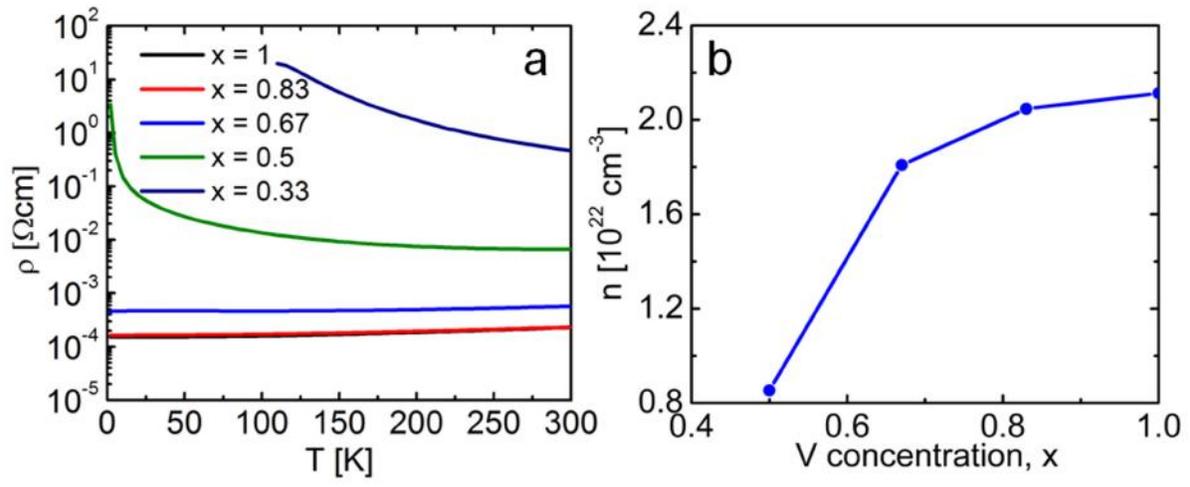

Fig. 3

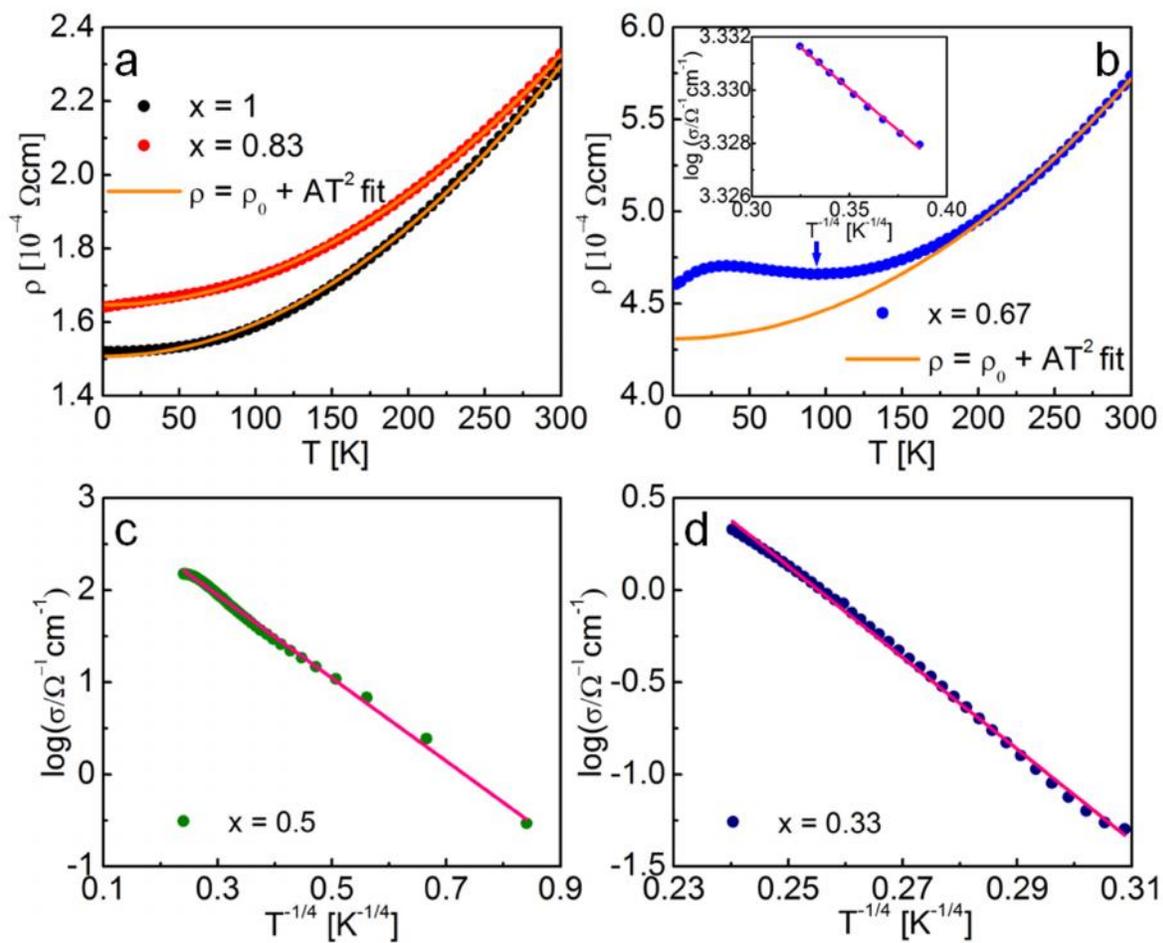

Fig. 4

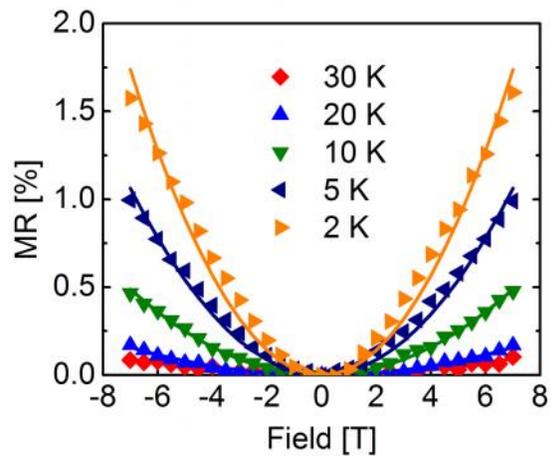